\documentclass{eptcs-modified}

\usepackage{amsmath}
\usepackage{amssymb}
\usepackage{url}
\usepackage{amsthm}
\usepackage{graphicx}
\usepackage{amscd}
\usepackage[usenames,dvipsnames,svgnames,table]{xcolor}

\begin{document}

\theoremstyle{definition}
\newtheorem{theorem}{Theorem}
\newtheorem{definition}[theorem]{Definition}
\newtheorem{problem}[theorem]{Problem}
\newtheorem{assumption}[theorem]{Assumption}
\newtheorem{corollary}[theorem]{Corollary}
\newtheorem{proposition}[theorem]{Proposition}
\newtheorem{example}[theorem]{Example}
\newtheorem{lemma}[theorem]{Lemma}
\newtheorem{observation}[theorem]{Observation}
\newtheorem{fact}[theorem]{Fact}
\newtheorem{question}[theorem]{Open Question}
\newtheorem{conjecture}[theorem]{Conjecture}
\newtheorem{addendum}[theorem]{Addendum}
\newcommand{\uint}{{[0, 1]}}
\newcommand{\Cantor}{{\{0,1\}^\mathbb{N}}}
\newcommand{\name}[1]{\textsc{#1}}
\newcommand{\id}{\textrm{id}}
\newcommand{\dom}{\operatorname{dom}}
\newcommand{\Dom}{\operatorname{Dom}}
\newcommand{\codom}{\operatorname{CDom}}
\newcommand{\spec}{\operatorname{spec}}
\newcommand{\opti}{\operatorname{Opti}}
\newcommand{\optis}{\operatorname{Opti}_s}
\newcommand{\Baire}{{\mathbb{N}^\mathbb{N}}}
\newcommand{\hide}[1]{}
\newcommand{\mto}{\rightrightarrows}
\newcommand{\Sierp}{Sierpi\'nski }
\newcommand{\BC}{\mathcal{B}}
\newcommand{\C}{\textrm{C}}
\newcommand{\CC}{\textrm{CC}}
\newcommand{\UC}{\textrm{UC}}
\newcommand{\lpo}{\textrm{LPO}}
\newcommand{\llpo}{\textrm{LLPO}}
\newcommand{\leqW}{\leq_{\textrm{W}}}
\newcommand{\leW}{<_{\textrm{W}}}
\newcommand{\equivW}{\equiv_{\textrm{W}}}
\newcommand{\equivT}{\equiv_{\textrm{T}}}
\newcommand{\geqW}{\geq_{\textrm{W}}}
\newcommand{\pipeW}{|_{\textrm{W}}}
\newcommand{\nleqW}{\nleq_\textrm{W}}
\newcommand{\leqsW}{\leq_{\textrm{sW}}}
\newcommand{\equivsW}{\equiv_{\textrm{sW}}}
\newcommand{\Sort}{\operatorname{Sort}}
\newcommand{\aouc}{\mathrm{AoUC}}
\newcommand{\pitc}{\Pi^0_2\textrm{C}}

\newcommand{\me}{P.~}

\title{An update on Weihrauch complexity, and some open questions}

\author{
Arno Pauly
\institute{Department of Computer Science\\Swansea University, Swansea, UK\\}
\email{Arno.M.Pauly@gmail.com}
}

\def\titlerunning{An update on Weihrauch complexity}
\def\authorrunning{A.~Pauly}
\maketitle

\begin{abstract}
I will give an update on progress in the study of Weihrauch degrees since the survey \cite{pauly-handbook} in 2017, and the Dagstuhl meeting in 2018 \cite{pauly-dagstuhl2}, and highlight some open questions. No claim of completeness or impartiality is made -- I would be grateful to be alerted about omissions, though.
\end{abstract}

\paragraph*{Algebraic structure of the Weihrauch degrees}
In logical terms, the algebraic structure of the Weihrauch degrees is $(\mathfrak{W}; 0,1,\top; \sqcap, \sqcup,\times,\star,\rightarrow,^*,\widehat{\phantom{f}},^\diamond)$, with the constants $0$, $1$, $\top$ being \emph{vacuously true}, \emph{true} and \emph{false} respectively. The $\sqcap$-operator is \emph{or}, and $\sqcap$, $\times$, $\star$ are all logical \emph{and}s, the additive, multiplicative and sequential \emph{and} respectively. The right-residual to $\star$ is $\rightarrow$, we do not have additional implications. The unary ``bang'' operators $^*$, $\widehat{\phantom{f}}$,$^\diamond$ provide access to finitely many parallel instances, countably-many parallel instances and finitely-many consecutive instances. A persistent open question on these operations was recently solved:

\begin{theorem}[Westrick \cite{westrick}]
$1 \leqW f$ and $f \equivW f \star f$, iff $f \equivW f^\diamond$.
\end{theorem}

\begin{question}
What can we say about $\mathrm{Th}(\mathfrak{W}; 0,1,\top; \sqcap, \sqcup,\times,\star,\rightarrow,^*,\widehat{\phantom{f}},^\diamond)$? Is its quantifier-free part finitely axiomatisable? What about fragments of the signature?
\end{question}

Another line of questions concerns definability of operators from others. As supremum and infimum, we know $\sqcup$ and $\sqcap$ to be definable in terms of $\leqW$. It is easy to see that $^*$ is definable from $1$, $\leqW$ and $\times$, in very much the same way that Westrick's result has shown $^\diamond$ to be definable from $1$, $\leqW$ and $\star$. By definition, $\rightarrow$ is definable from $\star$ and $\leqW$. What other definability results can we obtain? In particular, is $\widehat{\phantom{f}}$ definable from the rest of the structure? What about the jump of Weihrauch degrees (cf.~\cite{gherardi4})?

Brattka and Gherardi introduced the notion of completion of a Weihrauch problem, and showed that the parallelizable complete Weihrauch degrees form a Brouwer algebra \cite{gherardi2,gherardi6}. Goh has provided some clarifications on how the composition ought to be defined \cite{goh2}.

\paragraph*{Off the beaten track}
There is a canonic scaffold of familiar degrees in the Weihrauch lattice, the closed choice principles and the levels of the effective Baire hierarchy, which has served well for many classifications of computational problems. Recently however we have also studied some Weihrauch degrees that lie very much not in this explored part of the lattice. One such example is found in \emph{overt choice} for incomplete countably-based spaces. This principle is given overt (aka positive) information about a closed subset, and needs to output a point in this set.

\begin{theorem}[de Brecht, Schr\"oder \& P. \cite{paulydebrecht4}]
Let $\mathrm{VC}_\mathbb{Q}$ denote overt choice on the rationals. Then
\begin{enumerate}
\item If $f : \mathbf{X} \mto \mathbb{N}$ satisfies $f \leqW \mathrm{VC}_\mathbb{Q}$, then $f$ is computable.
\item $\mathrm{VC}_\mathbb{Q}$ is non-uniformly computable.
\item $\mathrm{VC}_\mathbb{Q} \nleqW \C_\mathbb{R}$
\end{enumerate}
\end{theorem}

For overt choice $\mathrm{VC}_\mathbf{X}$ for non-Fr\'echet-Urysohn coPolish spaces $\mathbf{X}$ we only know that $\mathrm{LPO} \leqW \mathrm{VC}_\mathbf{X} \leqW \widehat{\left (\operatorname{isEmpty} : \mathcal{A}(\Baire) \to \mathbb{S}\right )}$ -- which is perfectly compatible with $\mathrm{VC}_\mathbf{X}$ being a familiar Weihrauch degree, and is a remarkably weak classification.

\begin{question}[de Brecht, Schr\"oder \& \me \cite{paulydebrecht4}]
Can we classify overt choice for non-Fr\'echet-Urysohn coPolish spaces more precisely?
\end{question}

Another example for this is found in the principle $\mathrm{DS}$ that takes as input a linear order with a descending sequence, and then finds such a sequence. This principle sits, in a sense, beside the Baire hierarchy, as revealed by the following theorem:

\begin{theorem}[Goh, \me \& Valenti]
\begin{enumerate}
\item $\lim \leW \mathrm{DS} \leW \C_\Baire$
\item $\UC_\Baire \pipeW \mathrm{DS}$
\item If $f : \Baire \to \Baire$ is a function with $f \leqW \mathrm{DS}$, then $f \leqW \lim$
\end{enumerate}
\end{theorem}

\begin{question}
Does $\C_\Cantor' \leqW \mathrm{DS}$ hold?
\end{question}

\paragraph*{An either-or power of Ramsey's theorem}

Let $\mathrm{NON} : \Cantor \mto \Cantor$ be defined as $q \in \mathrm{NON}(p)$ iff $q$ is not computable relative to $p$. The following shows, in a sense, that Ramsey's theorem for pairs has both the power to force non-computable solutions, and to exhibit functional discontinuity -- but it cannot do both at once:

\begin{theorem}[Dzhafarov, Goh, Hirschfeldt, Patey \& P. \cite{pauly-patey}]
$\mathrm{NON} \times \lpo \nleqW \mathrm{RT}_2^2$
\end{theorem}

A negative answer to the following would also imply the celebrated separation $\mathrm{WKL} \nleqW \mathrm{RT}^2_2$ by Liu \cite{liu}.

\begin{question}
Does $\mathrm{NON} \times \llpo \leqW \mathrm{RT}^2_2$ hold?
\end{question}

\paragraph*{Belated closure under composition}
It may seem intuitive that natural Weihrauch degrees should either be closed under composition, or more and more iterations should give ever increasing power. We do know a few counterexamples to this. Let $\aouc$ be the degree of finding a solution to $bx = a$ where $0 \leq a \leq b$, and let $\mathrm{List}_{\Cantor}$ be the problem ``given a countable closed subset of Cantor space, find an enumeration of its elements''. We know:

\begin{theorem}[Kihara \& \me \cite{pauly-kihara2-mfcs}]
$\aouc^* \leW \aouc^* \star \aouc^* \equivW \aouc^\diamond$.
\end{theorem}

\begin{theorem}[Kihara, Marcone \& \me \cite{pauly-kihara4}]
$\mathrm{List}_{\Cantor} \leW \mathrm{List}_{\Cantor} \star \mathrm{List}_{\Cantor} \star \mathrm{List}_{\Cantor} \equivW \UC_\Baire$.
\end{theorem}

\begin{theorem}[Goh, \me \& Valenti]
$\mathrm{DS} \leW \mathrm{DS} \star \mathrm{DS} \equivW \C_\Baire$.
\end{theorem}

\begin{question}
Is there a square-root operator on the Weihrauch degrees?
\end{question}

\begin{question}
Does $\mathrm{List}_{\Cantor} \star \mathrm{List}_{\Cantor} \equivW \UC_\Baire$ hold?
\end{question}

\begin{question}
Is there any natural problem (e.g.~from linear algebra) in the Weihrauch degree $\aouc^\diamond$?
\end{question}

\paragraph*{Separation techniques}

To make proving the non-existence of reductions more feasible, techniques that instead let us prove positive results are very useful. One such technique is based on the recursion theorem:

\begin{theorem}[Kihara \& \me \cite{pauly-kihara5-arxiv}]
\label{theo:recursionweihrauch}
Let $\mathbf{X}$ have a total precomplete representation. Let $f : \mathbf{X} \mto \mathbf{Y}$ and $g : \mathbf{U} \mto \mathbf{V}$ be such that there exists a computable $e : \mathbf{U} \times \mathcal{M}(\mathbf{V},\mathbf{Y}) \mto \mathbf{X}$ such that if $x \in e(u,k)$ and $v \in g(u)$, then $k(v) \nsubseteq f(x)$. Then $f \nleqW g$.
\end{theorem}

\paragraph*{Analogues to $\mathrm{ATR}_0$}
Since its beginnings, Weihrauch complexity has drawn heavily on reverse math for inspiration and adaptable proofs. Analogues to four of the ``big five'' are readily available (although $\Pi^1_1-\mathrm{CA}$ was only recently considered by Hirst \cite{hirst5}), but the picture around $\mathrm{ATR}_0$ is more complicated. Recently, these part of the Weihrauch lattice has been explored by Kihara, Marcone and \me~\cite{pauly-kihara4}, by Angl\'es d'Auriac and Kihara \cite{dauriac}, by Goh \cite{goh} and by Marcone and Valenti \cite{valenti}. Some of the open questions from \cite{pauly-kihara4} were already answered in \cite{dauriac} and \cite{goh}, but the following remains open (see \cite{pauly-kihara4} for definitions):

\begin{question}
What is the relationship between $\mathrm{Det}_2$ and $\mathrm{PTT}_2$?
\end{question}

\paragraph*{Classifications of specific principles}
Many specific principles have been recently classified in the Weihrauch lattice, and there is no short supply of further challenges here. Dzhafarov, Flood, Solomon and Westrick classified some variants of the Dual Ramsey Theorem \cite{westrick2}. Marcone and Valenti have explored the open and clopen Ramsey theorems \cite{valenti}. Of the questions raised in \cite{valenti}, we draw attention to:

\begin{question}[{Marcone \& Valenti \cite[Question 6.2]{valenti}}]
Does $\Sigma^0_1\mathrm{-RT} \leqW \C_\Baire$ hold?
\end{question}

Davis, Hirschfeldt, Hirst, Pardo, \me and Yokoyama studied the principles that infinite sequences over a finite colour supply have an eventually repeating respectively and eventually constant tail regarding colour occurrence \cite{hirst-pauly}. Fiori-Carones, Shafer and Sold\`a studied the Rival-Sands theorem \cite{fiori}, and proved it to be equivalent to $\mathrm{WKL}''$. The operation of finding a projection of point to a closed (or overt) subset of $\mathbb{R}^n$ was classified by Gherardi, Marcone and \me \cite{pauly-marcone}. Depending on the dimension, and whether the set being projected to is given as closed, as overt or as both closed and overt set, various closed choice principles appear as degrees here. Kohlenbach has studied the Weihrauch degree of the modulus of regularity, used to establish convergence rates in convex optimisation \cite{kohlenbach2}. At the first Dagstuhl seminar on Weihrauch reducibility, Fouch\'e had asked about the Weihrauch degree of Fourier dimension (cf.~\cite{pauly-dagstuhl}). The answer was provided by Marcone and Valenti \cite{valenti2}. Crook and \me obtained some results on the degree of finding zeros of polynomials \cite{tonicha}.

\paragraph*{Connections to proof theory}
To what extent a Weihrauch reduction between two theorems can be related to the provability of one of the theorems from the other in a suitable proof system has been a focus of investigation for a while now, with Kuyper reporting initial results in \cite{kuyper}. Recent progress has been reported by Fujiwara \cite{fujiwara,fujiwara2}, Uftring \cite{uftring} and Hirst and Mummert \cite{hirst4}.

\paragraph*{Fine structure of the Weihrauch degrees}
Hertling \cite{hertling10} has extended the characterization of continuous Weihrauch reducibility for functions of certain types. Adapting this to also cover the usual Weihrauch reducibility seems a challenging task. Westrick pointed out that we lack even a clear picture of the Weihrauch reductions between functions with a single point of discontinuity (cf.~\cite{pauly-dagstuhl2}). A potentially related question was raised by Nobrega and \me:

\begin{question}[Nobrega \& \me \cite{pauly-nobrega}]
What is the Weihrauch degree of obtaining a point of discontinuity of a function from a Player I winning strategy in the Wadge game for it?
\end{question}

It is known that a computable winning strategy implies the existence of a computable point of discontinuity, but not whether there is a uniform argument to establish this.

\paragraph*{Connections to DST}
Day, Downey and Westrick showed that Bourgain's $\alpha$-rank can be characterized via Weihrauch reducibility \cite{westrick3}.

\paragraph*{Implementation in Coq}
Steinberg, Th{\'e}ry and Thies demonstrated the feasibility of doing computable analysis in Coq by implementing some sample results, including the Weihrauch separation $\C_\mathbb{N} \nleqW \id$ \cite{thies}.

\paragraph*{Work inspired by Weihrauch reducibiliity}
Greenberg, Kuyper and Turetsky \cite{greenberg2} explored an abstract scheme relating computability theory to the study of cardinal invariants inspired by Weihrauch reducibility. Carl explored using an OTM-variant of Weihrauch reducibility to compare the strength of axioms over $\mathrm{ZF}$ \cite{merlin}. Dzhafarov, Hirschfeldt and Reitzes considered a notion akin to Weihrauch reducibility relative to models of $\mathrm{RCA}_0$ (or other theories) \cite{reitzes}. Bauer and Yoshimura introduced the notion of instance reducibility, a special case of which are the extended Weihrauch degrees (which include representatives of false theorems, too) \cite{yoshimura3}.

\begin{question}
Is there a clear picture of how the additional degrees in the extended Weihrauch degrees relate to the Weihrauch degrees?
\end{question}

\begin{question}
What are some examples of interesting extended Weihrauch degrees of false statements, and how do they relate to the usual scaffolding Weihrauch degrees?
\end{question}

\bibliographystyle{eptcs}
\bibliography{../../spieltheorie}

\end{document}